\documentclass[twocolumn]{aastex62}

\usepackage{graphicx}
\usepackage{dcolumn}
\usepackage{bm}
\usepackage{hyperref}
\usepackage{cancel} 

\usepackage{blindtext}
\usepackage{braket}
\usepackage{booktabs}

\usepackage{wrapfig}
\usepackage{color}

\DeclareMathAlphabet\mathbfcal{OMS}{cmsy}{b}{n}
\DeclareMathAlphabet\mathcal{OMS}{cmsy}{n}{n}
\newcommand{\mb}[1]{\mathbf{#1}}

\def\aap{{Astronomy and Astrophys.}}

\def\apj{{\it Astrophys. J.\ }}
\def\apjs{{\apj\ \it Suppl.\ }}		
\def\apjl{{\apj\ \it Lett.\ }}

\submitjournal{ApJ}
\shorttitle{Elemental Selectivity in Cosmic Ray Acceleration}
\shortauthors{Hanusch et al.}
\begin{document}


\title{ACCELERATION OF COSMIC RAYS IN SUPERNOVA SHOCKS: ELEMENTAL SELECTIVITY OF THE INJECTION MECHANISM}

\correspondingauthor{Adrian Hanusch}
\email{adrian.hanusch@uni-rostock.de}

\author{Adrian Hanusch}%
\affiliation{%
 Institut f\"ur Physik, Universit\"at Rostock, 18051 Rostock, Germany
}%

\author{Tatyana V. Liseykina}%
\affiliation{%
 Institut f\"ur Physik, Universit\"at Rostock, 18051 Rostock, Germany
}%
\author{Mikhail Malkov}%
\affiliation{%
 CASS and Department of Physics, University of California, San Diego, La Jolla, California 92093, USA
}%

\date{\today}

\begin{abstract}

Precise measurements of galactic cosmic rays revealed a significant difference between the rigidity spectral indices of 
protons and helium ions. This finding is a notable contrast to the commonly accepted theoretical prediction that supernova 
remnant (SNR) shocks accelerate protons and helium ions with the same rigidity alike. Most of the earlier explanations for 
the "paradox" appealed to SNR environmental factors, such as inhomogeneous $p$/He mixes in the shock upstream medium, variable 
ionization states of He, or a multi-SNR origin of the observed spectra. The newest observations, however, are in tension with most 
of them.  In this paper, we show by self-consistent hybrid simulations that such special conditions are not vital for the explanation 
of the cosmic ray rigidity spectra. In particular, our simulations prove that an SNR shock can modify the chemical composition 
of accelerated cosmic rays by preferentially extracting them from a homogeneous background plasma without additional, largely 
untestable assumptions. Our results confirm the earlier theoretical predictions of how the efficiency of injection depends on the shock 
Mach number $M.$ Its increase with the charge-to-mass ratio saturates at a level that grows with $M.$ We have convolved the 
time-dependent injection rates of protons and helium ions, obtained from the simulations, with a decreasing shock strength over 
the active life of SNRs. The integrated SNR rigidity spectrum for $p$/He ratio compares well with the AMS-02 and PAMELA data.

\end{abstract}

\pacs{98.38.Mz, 98.70.Sa}
\keywords{cosmic rays, supernova remnants, acceleration of particles, shock waves}
                              

\section{Introduction}\label{sec:intro}
\begin{figure}[t]
  \includegraphics[width=0.95\columnwidth]{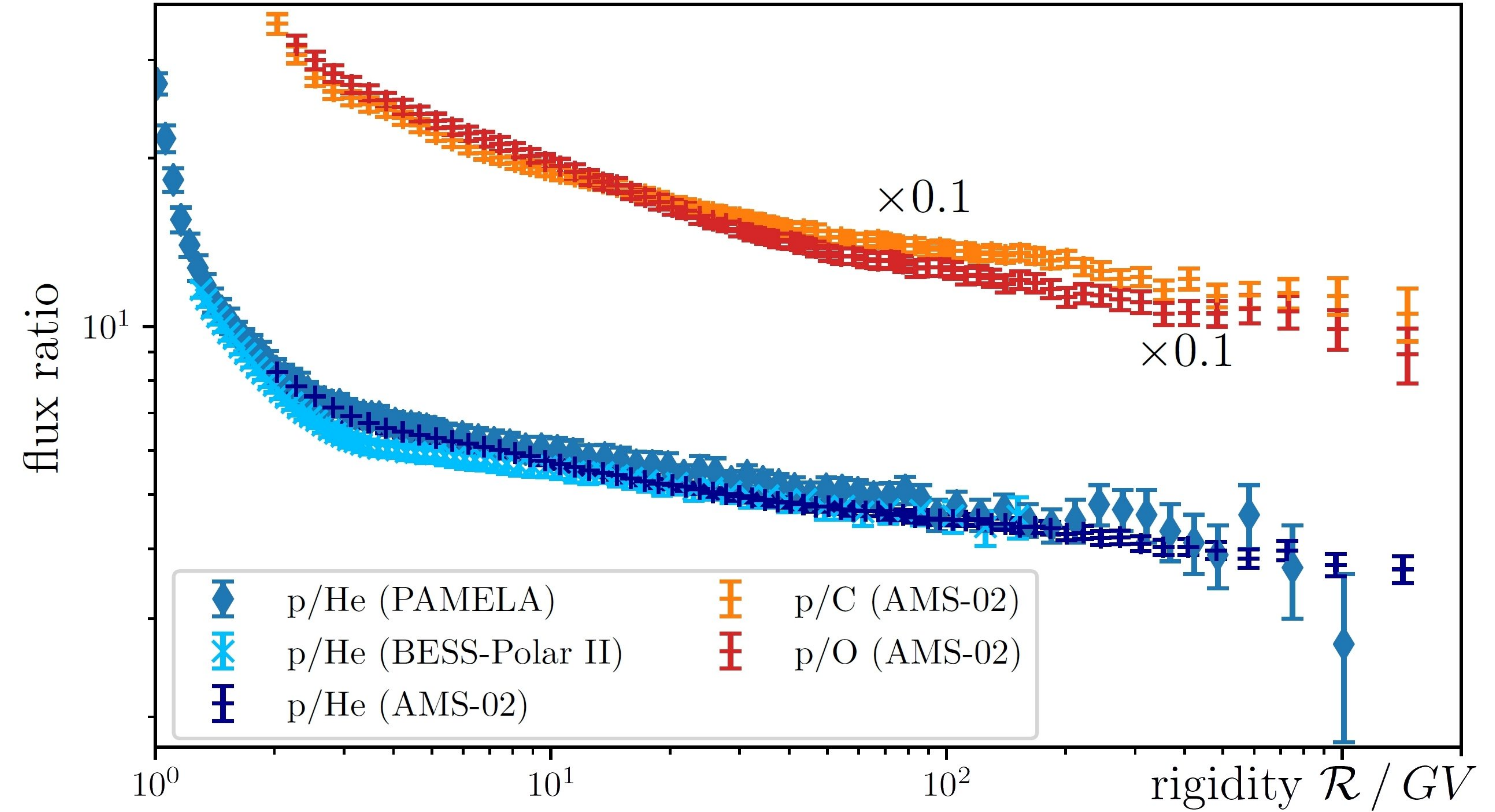}
  \caption{The $p$/He, $p$/C, and $p$/O ratio as a function of particle rigidity.
  The data is taken from \cite{Adriani11,Aguilar15,Abe16,Aguilar17}.
  \label{fig:The-p/He-ratio}}
\end{figure}

The PAMELA and AMS-02 measurements \citep{Aguilar15,Adriani11} indicated a 
difference $\Delta q\simeq 0.1$ between the rigidity spectral indices of protons and helium ions,
put forth earlier by the balloon-born experiment ATIC-2 \citep{Panov2009}.
According to 
observations \citep{Yoon11},
the scaling shown in Fig.~\ref{fig:The-p/He-ratio} is likely
to continue to higher rigidities. 
These findings challenge the hypothesis
of cosmic ray (CR) origin in the supernova remnants (SNR), see e.g., \citep{Bykov2018}, for a recent review. 

The leading CR production mechanisms, the first order Fermi or diffusive shock acceleration (DSA), is electromagnetic in nature
\citep{Fermi49}. 
The equations of motion of charged particles in arbitrary electric and magnetic fields
can be rewritten in terms of particle  
rigidity $\mathbfcal R=\mathbf{p}c/eZ$, instead of momentum $\mathbf{p}$:
\begin{equation}
  \frac{1}{c}\frac{d\mathbfcal R}{dt} 
  = \mathbf{E}\left(\mathbf{r},t\right) 
  + \frac{\mathbfcal R\times\mathbf{B}\left(\mathbf{r},t\right)}{\sqrt{\mathcal{R}_{0}^{2}+\mathcal{R}^{2}}},
  \label{eq:RigMotion}
\end{equation}
\begin{equation}
  \frac{1}{c}\frac{d\mathbf{r}}{dt} 
  = \frac{\mathbfcal R}{\sqrt{\mathcal{R}_{0}^{2}+\mathcal{R}^{2}}}.
  \label{eq:CoordMotion}
\end{equation}
Here $\mathcal{R}_{0}=Am_{p}c^{2}/Ze$, with $A$ being the atomic
number and $m_{p}$ the proton mass. The electric, $\mathbf{E}\left(\mathbf{r},t\right)$, and magnetic,
$\mathbf{B}\left(\mathbf{r},t\right)$, fields here are completely
arbitrary. So, the equations apply not only to the acceleration of
CRs in a SNR shock but also to their propagation through the turbulent
interstellar medium (ISM) to an observer. Moreover, the propagation includes an eventual escape of the accelerated CRs from the Milky Way.
The  equations (\ref{eq:RigMotion}) and (\ref{eq:CoordMotion}) show that all species with rigidities $\mathcal{R}\gg\mathcal{R}_{0}=Am_{p}c^{2}/Ze$ 
have nearly identical orbits in the phase space $\left({\bf r},\mathbfcal R\right)$. Hence,  
if different elements enter the acceleration in a {\emph{time-independent}} ratio 
at some $\mathcal{R}\gg\mathcal{R}_{0}=Am_{p}c^{2}/Ze,$ their 
rigidity spectra in this range should be identical \citep{MalkovPRL12,MalkovArXiv17}, in apparent contradiction with \mbox{ATIC-2}, PAMELA, AMS-02 observations, Fig.~\ref{fig:The-p/He-ratio}. 
Note, that at reasonably low rigidities, such as 10 GV and lower, where also solar modulation is observed, the
rule of equal rigidity argument does not apply. The reason is that the rest-mass rigidity, $\mathcal{R}_0 \approx 1$ GV
for protons, does enter the equations of motion.

To explain the $p$/He rigidity paradox three ideas have been entertained: 
 (1) shock {\emph {evolution in time}}; (2) contributions from several SNRs with different $p$-He mixes and
spectral slopes; (3) CR spallation in the ISM that introduces particle
sources and sinks in their kinetic equations.
Turning to the first idea, assume that the $p$/He ratio is known at some fiducial rigidity $\mathcal{R}=\mathcal{R}_{1}\gg\mathcal{R}_{0}.$
While the shock strength naturally decreases, this ratio must increase at a rate consistent with the 
observed $p$/He slope in rigidity. The crucial point here is that the power-law index of shock-accelerated 
particles decreases with the shock Mach number $M(t)$ rather definitively
\begin{equation}
  q = - \frac{d \ln f}{d \ln \mathcal{R}} = \frac{4}{1-M^{-2}},
  \label{eq:qOfM}
\end{equation}
where $f$ is the CR distribution function. 
Therefore, to produce
a $p$/He fixed index at $\mathcal{R}>\mathcal{R}_{1}$, the $p$/He
ratio at $\mathcal{R}=\mathcal{R}_{1}$ must depend on $M(t)$ in a specific way. 
If this dependence is an intrinsic property of collisionless shock, it cannot
be adjusted to fit the data, thus making the scenario (1) fully testable.
Unlike the scenario 
(1) above, 
(2) is not testable because the individual
properties of contributing sources are unknown.
Besides, it will likely
fail the Occam's razor test, especially after the AMS-02 has measured
$p$/C and $p$/O ratios to be identical to those of $p$/He \citep{Aguilar17}.
And moreover, it would be impossible to maintain the spectral slopes in the ratios $p$/He, $p$/C and 
$p$/O \citep{Aguilar17,Aguilar18} nearly constant over an extended rigidity range \citep{MalkovArXiv17}. 
As for the spallation effects (3), the equivalence between the He, C and O spectra \citep{Aguilar18} 
corroborates the conclusion \citep{Vladimirov12}  that it is 
insufficient to explain the observed differences between $p$ and elements 
whose $A/Z$ values are similar but higher than that of the protons. 
It follows that the time dependence of the subrelativistic 
acceleration phase, i.e. injection into DSA, 
option (1), is the most realistic scenario to consider. 

Time dependence of particle acceleration at an SNR shock comes in two flavors. Firstly,
the natural shock weakening 
makes the acceleration time dependent. Secondly, the medium into which the shock 
propagates may be inhomogeneous (effect of SNR environment) \citep{Ohira11}. 
If also the background $p$/He ratio is inhomogeneous and increases outward, after 
the acceleration it will decrease with rigidity. This is because higher rigidities are 
dominated by earlier times of acceleration history when the He contribution was 
higher.  
The problem with this explanation is that
not only the He concentration must decrease with
growing shock radius at a specific rate (one free parameter), 
but  so must C and O. This conclusion follows from the newest
C/He and O/He AMS-02 flux ratios, which have turned out to be independent of rigidity \citep{Aguilar17,Aguilar18}.
So, He, C, and O are likely to share their 
acceleration and propagation history. One natural consequence of this is that C and O are unlikely 
to be preaccelerated from grains, contrary to some earlier suggestions (see \citep{Ohira16} for the recent 
study and earlier references).
Note that it is crucial to use the rigidity dependence of the \emph{fractions} 
of different species as a primary probe into the intrinsic properties
of CR accelerators. Unlike the individual spectra, the fractions are
unaffected by the CR propagation, reacceleration, and losses from
the galaxy, as long as spallation is negligible. 

Besides tensions with the recent AMS-02 results, the above-discussed mechanisms
require additional and untestable assumptions. 
To resolve these problems,  
\citep{MalkovPRE98}  argued that in quasi-parallel
shocks a specific elemental selectivity of the initial phase of the DSA (injection) 
occurs with no additional assumptions. Analytic calculations 
have established that the ion injection efficiency into the DSA depends on the shock Mach number,
increases with $A/Z$, and saturates at a level that
grows with $M$. 
The publication of measurements of 
$p$/He ratio by the
PAMELA collaboration \citep{Adriani11}, prompted the authors of \citep{MalkovPRL12}
to apply the analytic injection theory to the case $A/Z=2$ (specifically
to He$^{2+}$, also valid for fully stripped C and O, accurately measured
later by AMS-02), producing  
an excellent fit to
the PAMELA data in the relevant rigidity range $2<\mathcal{R}<200$~GV.
Moreover, the analytic results are largely insensitive to the ionization multiplicity 
at higher A/Z since the saturation effects becomes significant already at $A/Z\sim 2-4$ (see Fig.5 in \citep{MalkovPRE98}).  
Note that lower rigidities are strongly affected by solar modulation,
while at higher rigidities the PAMELA statistics was insufficient
to make a meaningful comparison.

In this paper we demonstrate that the recent high-precision
measurements of elemental spectra with different $A/Z$ are not only
consistent with the hypothesis of CR origin in the SNR, but also strongly
support it. Although a similar stand has been taken in \citep{MalkovPRL12}
about the PAMELA findings \citep{Adriani11}, the new AMS-02 data \citep{Aguilar17,Aguilar18} and
recent progress in shock simulations allow us to establish crucial
missing links in the CR-SNR relation. In particular, the coincidence
in accelerated particle spectral slopes of three different elements
with $A/Z\simeq2$ (He, C, and O) discovered by the AMS-02 experiment
points to an intrinsic, $A/Z$-based selection mechanism
and rules out incidental ones, such as particle injection from inhomogeneous
shock environments, preacceleration of elements locked into grains, 
or a variable ionization state of He \citep{Serpico2015}. 
It is important to emphasize here
that the latter mechanism was primarily justified by an integrated
abundance of different elements, whereas the detailed rigidity spectra
have become known only now.  
On the theoretical side, the $p$/He calculations \citep{MalkovPRL12}
are based on an analytic theory \citep{MalkovAA95} that 
allows freedom  in selecting
seed particles for injection. Pre-energized particles evaporating from the shocked downstream plasma back 
upstream \citep{ParkerJNE61,Quest88} and shock reflected
particles \citep{Burgess_Scholer2015}
have been most often discussed. 
Simulations can remove this uncertainty, 
thus greatly improving
the understanding of the $A/Z$ selectivity mechanism. 

Suprathermal protons (shock-reflected, or "evaporating"  
from hot downstream plasma) drive unstable Alfv\'en waves in front of the shock.
These waves control the injection of all particles by regulating their
access to those parts of the phase space from where they can repeatedly
cross the shock, thus gaining more energy \citep{KatoApJ2015,Markowith2016}. 
Furthermore, the waves are almost frozen into the local fluid. So, when crossing
the shock interface, they trap most particles and prevent them from
escaping upstream $again$, thus significantly reducing their odds for injection.
As protons drive these waves, the waves also trap protons most efficiently,
while, e.g., He$^{2+}$ have somewhat better chances to escape from the proton-generated waves 
upstream and to get eventually injected. The trapping becomes
naturally stronger with growing wave amplitude, that also grows with the Mach number. 
This trend is more pronounced for protons than for
He ions, which is crucial for the injection selectivity. 

Simulations remove another potentially important
limitation of the 
analytic treatment \citep{MalkovPRL12}. Namely, He ions
have not been included in the wave generation upstream and treated
only as test-particles. Such approximation is often considered to be sufficient
because of the large, $\simeq10,$ $p$/He number density ratio. However,
the He ions drive resonant waves that are typically two times longer than the
waves driven by the protons. In the wave-particle interaction, the
resonance condition is often more important than the wave amplitude.
In addition, the rational relation between the respective wave lengths
is suggestive of parametric interactions between them. Such
interaction should facilitate a cascade to longer waves which are
vital for the DSA, not just for particle injection.
Several hybrid simulations, addressing the
acceleration efficiency of alpha particles, did include them self-consistently
\citep{Scholer1998,Scholer1999,BurgessGRL89,TrattnerGRL91}, however in some cases with \emph{dramatically} reduced 
abundances 
\citep{CaprioliPRL17}, making them dynamically unimportant and thus completely excluding 
the He-driven waves. In Section \ref{sec:He-driven} we further discuss the influence of these waves 
on energy spectra of proton and He ions.
Besides, the earlier fully self-consistent  simulations, facing the problem of injection of different ions \citep{CaprioliPRL17}, did not provide sufficiently
detailed Mach number scans of the $p$/He injection ratio, that is
needed to test the theoretically predicted $p$/He injection bias.

\section{Simulation set-up}\label{sec:Set_up}
The full kinetic modeling of ion injection for a realistic ion-to-electron mass ratio $m_i/m_e$ 
is challenging because of the necessity to resolve both the electron- and ion-scales.
In this paper we study the particle injection into the DSA using hybrid simulations \citep[and references therein]{LipatovBook},
where only the ion plasma population is treated kinetically, while electrons are treated as a charge 
neutralizing massless fluid. The hybrid simulations have been proven to be a powerful tool in the investigation of the 
non-relativistic shocks and have been used for a variety of problems \citep[and references therein]{LipatovBook}, 
including the injection of protons into the DSA process \citep{Caprioli2015} and the study of the magnetic turbulence 
driven by plasma instabilities \citep{Caprioli2014}.
The underlying equations and implementation details are documented in the Appendix. 
The electron pressure $p_{e}$ and the resistivity are both assumed to
be isotropic quantities. The pressure $p_{e}$ is modeled
using an adiabatic equation of state with the adiabatic index
$\gamma_{e}=5/3.$ 
The fluid equations and the ion equations of motion are
non-relativistic, as $|\mathbf{v}| \ll c$ holds during the injection phase. 
In the simulations, lengths are given in units of $c/\omega_{p}$,
with $\omega_{p}=\sqrt{4\pi\,n_{0}\,e^{2}/m_{p}}$ being the proton
plasma frequency, $n_{0}$ the upstream density and $e$ the proton charge.
Time is measured in the units of inverse proton gyrofrequency, $\omega_{c}^{-1}=(e\,B_{0}/m_{p}\,c)^{-1}$. Here $B_{0}$ is the magnitude
of the background magnetic field.

We use a realistic composition of the plasma consisting of 
ion species with number ratios corresponding to the amount of particles in the ISM. 
The fraction of ions respective to protons is $\sim 10 \%$ for helium 
and $\sim 0.04\%$ for carbon and oxygen.
Note, that He ions are dynamically important and cannot be regarded as 
test particles.
The simulations are 1D in space but 3D in velocity and field components. 
This setting substantially increases the particle 
statistics and grid resolution,  lowers the noise, and improves wave 
description, all being crucial for understanding the downstream thermalization. 
It is more important than possible shock rippling effects, 
not captured by 1D simulations. 
Besides that, shock rippling and its impact on particle reflection cannot be accurately characterized within hybrid simulations
and require a full kinetic treatment \citep{Liseykina15,MalkovPoP16,MalkovArXiv17}. 
For the reasons explained in detail in sections \ref{sec:2Dvs1D} and \ref{sec:He-driven} we deliberately choose the 1D 
treatment as a first step in a systematic study of the $A/Z$ dependence of injection efficiency. The follow up 2D simulations will be discussed elsewhere.

The simulation is initiated by sending a supersonic and superalfv\'enic
plasma flow with velocity $v_{0}$
against a reflecting
wall. The shock forms due to the interaction of the emerging counter-propagating
flows. The background magnetic field is set 
parallel to the shock
normal $\mathbf{B}_{0}=B_{0}\mathbf{x}$. 
The upstream plasma betas
are $\beta_{e}=\beta_{i}=1$. The simulation box has a length 
of  $12 - 48 \cdot 10^3\,c/\omega_{p}$, depending on the initial velocity $v_{0}$. 
The spatial resolution of $\Delta x=0.25\;c/\omega_{p},$ 100 particles per species 
per numerical cell are used. The time step is 
$\Delta t=0.01/(v_{0}/v_{A})\;\omega_{c}^{-1}$ with $v_{A}=B_{0}/\sqrt{4\pi\,n_{0}\,m_{p}}.$ 
All numerical
parameters have been checked for convergence.\\

\section{Simulation results}\label{sec:Results}

\subsection{$A/Z$ dependence of injection}

\begin{figure}[t]
  \includegraphics[width=\columnwidth]{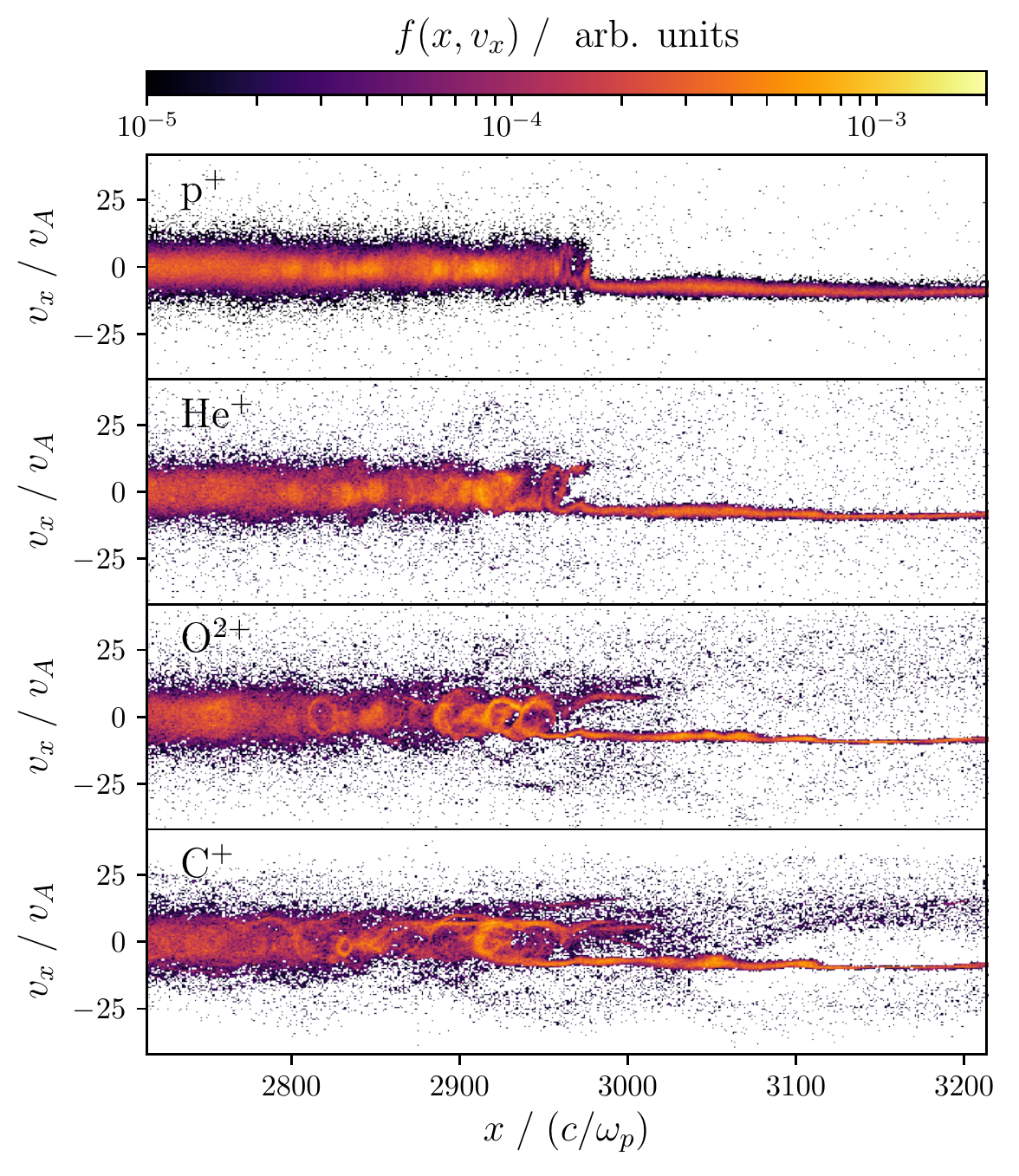}
  \caption{Phase space $f(x, v_x)$ for different ion species at 
    $t=1000 \; \omega_c^{-1}$. The plots show only a part of the 
    simulation domain and are centered around the shock transition.
  \label{fig:phase_space}}
\end{figure}

We investigate the mass-to-charge  dependence of the injection 
using self-consistent simulations for ion species with $A/Z\le 16.$
In addition to protons, 
He, C, and O ions with charge states $Z=1$ and $Z=2$ were included. 
The phase space distributions of selected ion species are shown in 
Fig.~\ref{fig:phase_space} for an upstream flow velocity $v_0 = 10 \, v_A.$ 
The transition from the cold upstream flow to the hot and turbulent downstream 
plasma is clearly seen in the plots.
The width of the particle distribution in $v_x$ in the downstream is almost the same for all
ion species, indicating higher temperatures for heavier species. 
The latter also thermalize further downstream 
as their impact from the shock is smoother.
The presence of ions with large $|v_x|$ in the up- and downstream
shows that some ions have already gained energy and are able to
cross the shock front.

\begin{figure}[t]
  \includegraphics[width=\columnwidth]{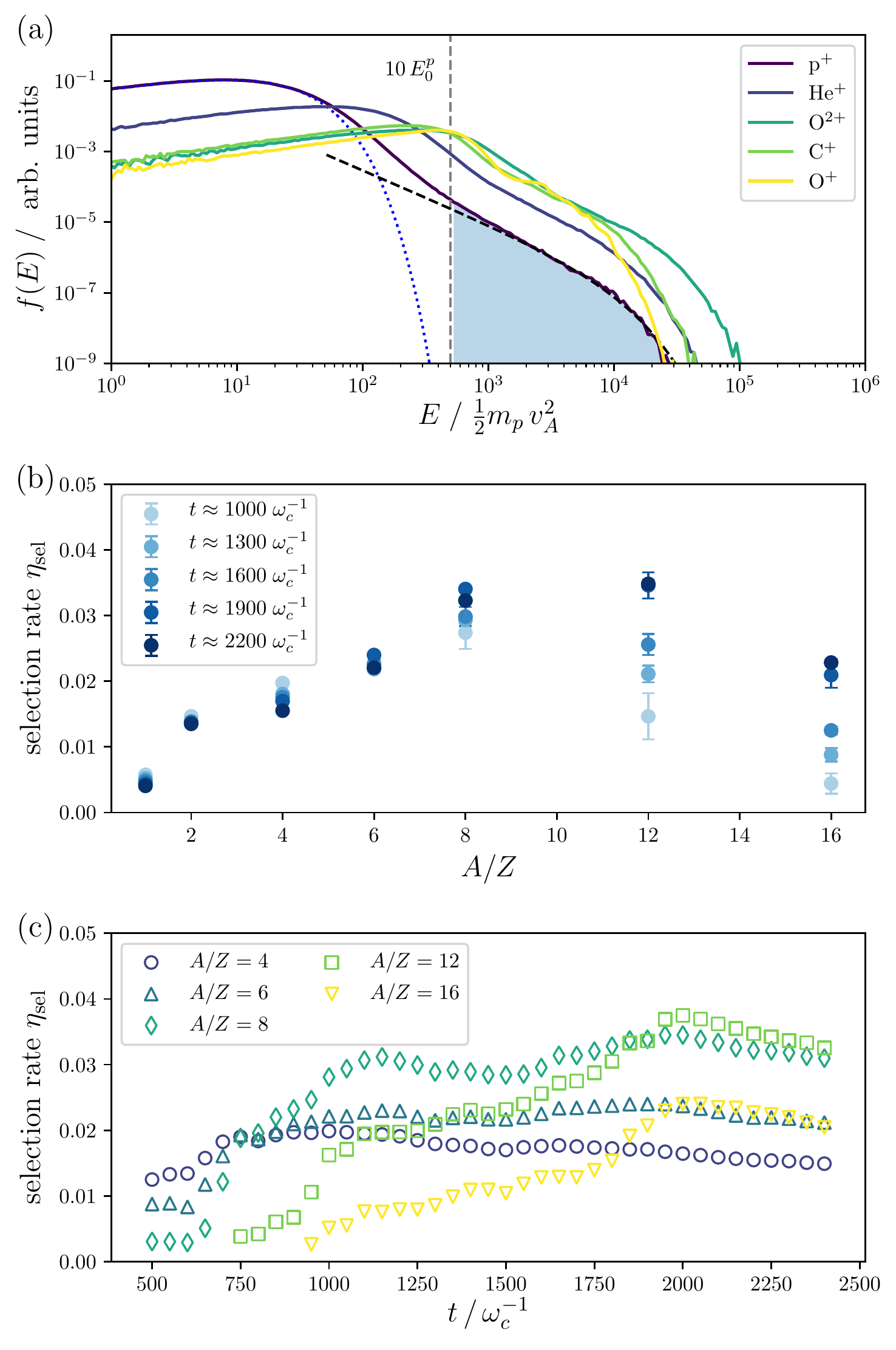}
  \caption{(a) Downstream energy spectra at $t=1500 \, \omega_c^{-1}$ for selected ion species present in
    the simulation. The protons are 
    in the tail of the distribution 
    function if their energy exceeds $10 \, E_0^p$ (dashed gray line).  The shaded area denotes the part of the spectrum, used for calculating the selection rate. (b) Selection rate $\eta_{\mathrm{sel}}$  
    as function of the mass-to-charge ratio for $v_0 = 10 \; v_A$, measured at different times between $t=1000 \, \omega_c^{-1}$ and $t=2200 \, \omega_c^{-1}$.
    (c) Selection rate as a function of time for selected ion species. The measurement of $\eta_{\mathrm{sel}}$ for heavy species with $A/Z> 6$ for $t<1000  \, \omega_c^{-1}$ is 
    not precise, because at this time the power tail of the distribution function is not yet well developed. Temporal evolutions $\eta_{\mathrm{sel}} (t)$ 
    shows how  the time at which $\eta_{\mathrm{sel}}$ saturated depends on A/Z.
  \label{fig:number_efficiency}}
\end{figure}

The energy spectra of the particles downstream of the shock
transition at $t = 1500 \; \omega_c^{-1}$, obtained using a logarithmic binning procedure, 
are shown in Fig.~\ref{fig:number_efficiency}a. 
The spectra of all ion species exhibit two main features: Maxwellian
distribution and a power-law tail. 
The transition from the Maxwellian to the power-law tail is obscured 
by a contribution of suprathermal particles \citep{Burgess_Scholer2015}.
The spectra of heavier ions are shifted to higher energies, 
as the velocity is randomized during the shock crossing.
For all species, $\alpha$, the tail is clearly developed for energies 
$E > 10\, E_0^{\alpha}$ with $E_0^\alpha = \frac{1}{2}\, m_\alpha\, v_0^2$. 
This energy is marked in the proton energy spectrum in 
Fig.~\ref{fig:number_efficiency}a by the dashed gray line. 
After the spectra are converged ($t \ge 2000\; \omega_c^{-1}$ for $v_0 = 10 \; v_A$), we
calculate the selection rate, $\eta_{\mathrm{sel}}$, i.e., the fraction of particles in the tail of the distribution function, as a function of mass-to-charge-ratio $A/Z.$
For low $A/Z,$ Fig.~\ref{fig:number_efficiency}b, $\eta_{\mathrm{sel}}$ grows almost linearly, 
a saturation occurs around $A/Z \sim 8-12$ (in a \emph{Mach-dependent fashion}, though) 
and at higher $A/Z$ 
the selection rate decreases, recovering a physically correct $A/Z\to\infty$ asymptotic behavior (it should tend to zero, as for the injection of neutrals). 
The preliminary 2D simulations also evidence a deviation from the linear $\eta_{\mathrm{sel}} (A/Z)$ trend, pointing towards a saturation for higher $A/Z.$
Note, that the exact position of the saturation is to some extent time-dependent as heavier ions are accelerated at later times, Fig.~\ref{fig:number_efficiency}c.
This is because the respectively longer waves need to be generated by the increasing maximum energy of protons.
The efficiency of these waves for injection of higher $A/Z$ species naturally depends on their amplitudes. These amplitudes depend not only on 
the maximum momentum of (resonant) protons but also on the dynamics of the entire wave spectrum, i.e., spectral transfer rate, turbulent cascade, etc. 
Our $\eta_{\mathrm{sel}}(A/Z)$ scaling (Fig.~\ref{fig:number_efficiency}b) is in agreement up to $A/Z\le 8$ with an almost linear increase of the selection rate 
 with mass-to-charge ratio found recently in 2D hybrid simulations, facing the problem of injection of different ions in quasi-parallel 
shocks  with $M>5$ ~\citep{CaprioliPRL17}. 
For more extensive discussion about the behavior of $\eta_{\mathrm{sel}}$ for high values of $A/Z,$ see Section \ref{sec:trend}.

\subsection{Elemental selectivity: proton-to-helium ratio}
\begin{figure}[t]
  \includegraphics[width=\columnwidth]{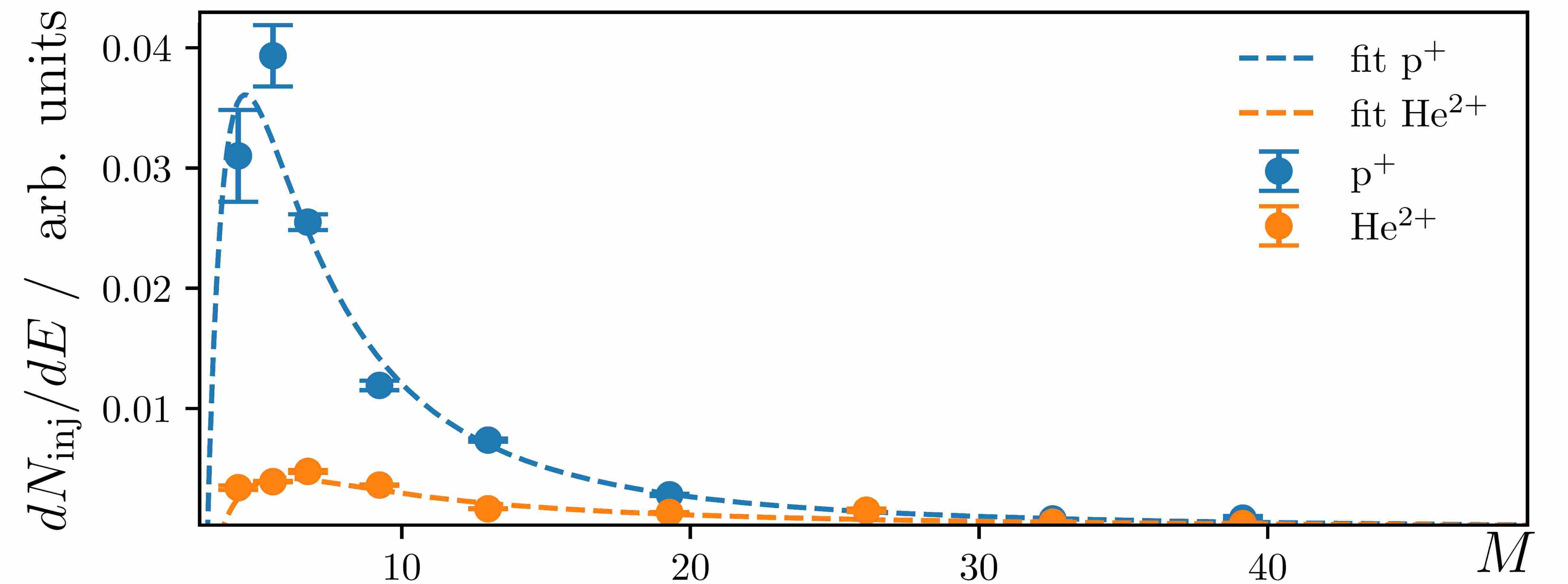}
  \caption{ Injection efficiencies of protons and He$^{2+}$  
  obtained from the simulation according to Eq.~(\ref{eq:inj-efficiency})
  as a function of Mach number. The dashed lines are the fits $\eta_{\mathrm{inj}}(M)=a\,(M-b)\,M^{-c},$
  with parameters $a_p=5.68, b_p=3.27, c_p=3.50$,
  and $a_\mathrm{He}=0.25, b_\mathrm{He}=3.79,~c_\mathrm{He}=2.73.$ 
  \label{fig:inj-efficiency}}
\end{figure}

 In the following we investigate the elemental selectivity of the injection  by focusing on the $p$/He ratio. To extract this quantity we calculate the injection 
efficiency of $p$ and He$^{2+}$ separately.
The direct measurement of the injection efficiency is difficult, because 
the transition from the Maxwellian distribution to the power-law tail is not sharp. 
Therefore, we fit a thermal distribution 
$f_{\mathrm{th}} \propto E^{1/2}\,\exp(-E/T)$ 
as well as a power-law with a cut-off, 
$f_{\mathrm{pow}} \propto E^{-q}\,\exp(-E/E_{\mathrm{cut}})$
to the low and high energy parts of the downstream spectrum.
Here $T$ is the downstream temperature of the respective ion species
and $E_{\mathrm{cut}}$ is the cut-off energy. 
In Fig.~\ref{fig:number_efficiency}a 
the dotted 
line denotes the fitted Maxwellian, while the dashed 
line is the power-law fit to the energy spectrum of protons. 
With $E_{\mathrm{inj}}$ defined for each species from $f_{\mathrm{th}}(E_{\mathrm{inj}}) 
= f_{\mathrm{pow}}(E_{\mathrm{inj}})$ the injection 
efficiency is calculated
as
\begin{equation}
 \eta_{\mathrm{inj}} \propto \left.\left(\frac{dN}{dE}\right)\right|_{E=E_{\mathrm{inj}}}= \frac{f_{\mathrm{th}}(E_{\mathrm{inj}})}{\int_{0}^{\infty}f_{\mathrm{th}}(E)\;\mathrm{d}E}.
  \label{eq:inj-efficiency}
\end{equation}
Figure~\ref{fig:inj-efficiency} shows the value of  $\eta_{\mathrm{inj}}^{\alpha}(M)$
obtained from a series of simulations with different
initial upstream flow velocities $v_{0}.$ The corresponding Alfv\'enic shock Mach numbers are 
$M =(v_{0}+v_{s})/v_{A}. $ Here $v_{s}$ is  the 
shock velocity in the downstream rest frame. 
The $M$- dependence of $\eta_{inj}$
is similar for $p$ and He$^{2+}$. It increases
for $M\lesssim5$ for protons, $M\lesssim7$ for He ions and 
decreases at higher $M$, tending to the predicted \citep{MalkovPRE98}  $\eta_{\mathrm{inj}}(M)\sim\ln M/M$ asymptotics.
  
Two aspects are important here. First, the injection of protons dominates
for low $M$ with $\eta_{\mathrm{inj}}^{p}$ exceeding
the value of $\eta_{\mathrm{inj}}^{\mathrm{He}}$ by an order of magnitude. Second, the maximum of $\eta_{\mathrm{inj}}^{p}$ is shifted towards smaller $M$ compared to He$^{2+}$. The prevalence of proton injection at weak shocks is also noticeable in the downstream temperature ratio
$T_{\mathrm{He}}/T_{p}$, which for $M<15$ exceeds the expected ratio of $T_{\mathrm{He}}/T_{p}=4.$

\subsection{Rigidity spectra}  To model the time-dependent CR acceleration we combine the Mach number dependent injection efficiency obtained from simulations, 
Fig.~\ref{fig:inj-efficiency},
with the theoretical spectral slope, $q=4/(1-M^{-2})$ that
allows us to extend the simulation
spectra far beyond in rigidity that any simulation may possibly reach.
The extension is justified by simulation spectra
reaching the asymptotic DSA power-law, Fig.~\ref{fig:number_efficiency}a.

During Sedov-Taylor phase of the SNR evolution
the shock radius increases with time as $R_{s}\simeq C_{\mathrm{ST}}\,t^{2/5}$,
while the shock velocity decreases as $V_{s}\simeq(2/5)\,C_{\mathrm{ST}}\,t^{-3/5},$ 
 with $C_{\mathrm{ST}}\simeq(2\,E_{e}/\rho_{0})^{1/5}$. Here $E_{e}$
is the ejecta energy of the supernova, $\rho_{0}$ is the ambient density. 
The number of CR species $\alpha,$ deposited in the shock interior, 
as the shock radius increases from $R_{\mathrm{min}}$ to $R_{\mathrm{max}}$, amounts to
\begin{equation}
  N_{\alpha}(p) 
  \propto \!\! \int\limits_{R_{\mathrm{min}}}^{R_{\mathrm{max}}} \!\! f_{\alpha}\left(\mathcal{R},M(R)\right)R^{2}\,\mathrm{d}R
  \propto \!\! \int\limits_{M_{\mathrm{max}}^{-2}}^{M_{\mathrm{min}}^{-2}} \!\! f_{\alpha}(\mathcal{R},M)\,\mathrm{d}M^{-2}.
  \label{eq:number}
\end{equation}

\noindent The spectra $f_{\alpha}$ are  

\begin{equation}
      f_{\alpha}\propto\eta_\mathrm{inj}^{\alpha}(M)\left({\mathcal{R}}/{\mathcal{R}_{\mathrm{inj}}}\right)^{-q(M)} 
  \label{eq:spectra}
\end{equation}
with $\displaystyle q(M)=4/(1-M^{-2}).$ 
Eqs.~(\ref{eq:number}) and (\ref{eq:spectra}) are accurate
for the most interesting sub-TV particles that are accelerated quickly.
Eq.~(\ref{eq:number}), however, tacitly imply an unimpeded release of accelerated
particles into the ISM which is poorly known. The key to our approach is that $p$/He ratio is still independent of the 
release mechanism and even ensuing propagation across the ISM, simply because the underlying equations of motion are identical 
for $p$ and He.

\begin{figure}[t]
  \includegraphics[width=\columnwidth]{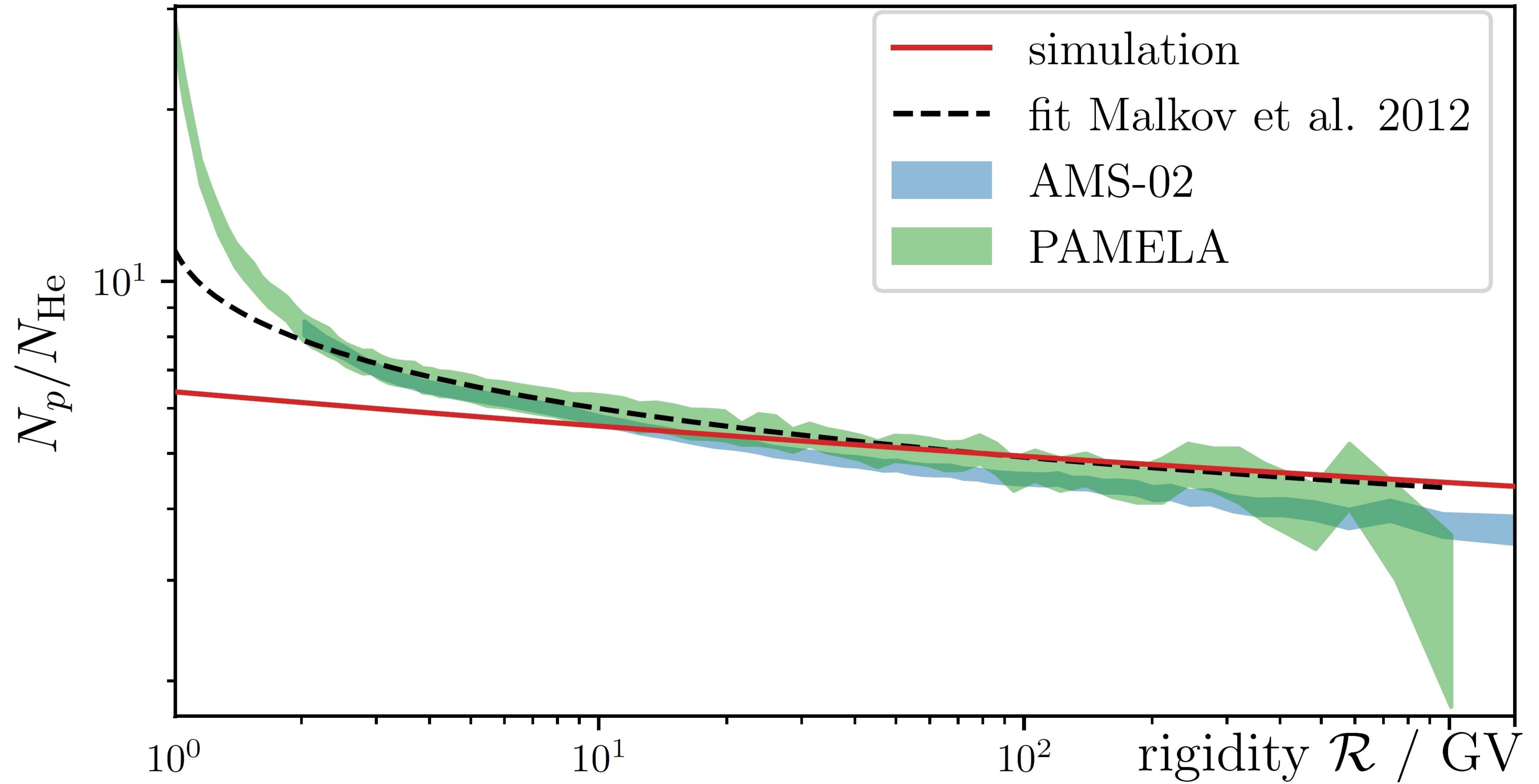}
  \caption{Proton-to-helium ratio as a function of particle rigidity. The results
  from the simulation (red line) are compared to the  PAMELA and AMS-02
  data. For details of the fit (dashed line) see Fig.~3 in \cite{MalkovPRL12}.
  The observed $p/$He ratio is accurately reproduced in the range 
  $\mathcal{R}\gtrsim10$ GV.
  \label{fig:pHe-ratio}}
\end{figure}%
Instead of feeding the simulation data for $\eta_\mathrm{inj}^{\alpha}\left(M\right)$
to the convolution given by Eqs.~(\ref{eq:number}) and (\ref{eq:spectra})
we first fit the following simple
function $\eta_{\mathrm{inj}}(M)=a\,(M-b)\,M^{-c}$  in the range $M_\mathrm{min} = 3.5<M<M_\mathrm{max} = 100$ to the data extracted
from the simulations, Fig.~\ref{fig:inj-efficiency}, and then calculate the $p$/He ratio,
$N_{p}/N_{\mathrm{He}}$, according to Eq.~(\ref{eq:number}), as
a function of rigidity. 
The resulting  $p$/He spectrum (red line), shown in Fig.~\ref{fig:pHe-ratio}, compares well in the high-rigidity range with the
AMS-02 and PAMELA data (shadow areas).

\section{Discussion}\label{sec:discussion}
\subsection{$A/Z$ trend of the selection rate}\label{sec:trend}
Our simulations show, Fig.\ref{fig:number_efficiency}~(b), that selection rate $\eta_{\mathrm{sel}}$ growth with mass-to-charge ratio, 
saturates in a Mach-dependent fashion around $A/Z\sim 8-12$, and then decreases for higher $A/Z$ values, recovering a physically correct $A/Z\to\infty$ asymptotic behavior, 
as expected for neutral particles. However, this result is in contradiction with the findings recently reported in \citep{CaprioliPRL17}, where the authors  
have obtained in 2D hybrid simulations the \emph{quadratic} growth with $A/Z$ of the chemical enhancement for mass-to-charge ratios as high as $A/Z = 56.$ 
The striking contradiction consists in the fact, that the quadratic growth of the chemical enhancement implies the \emph{linear} growth 
of the selection rate with $A/Z$ \emph{without trend to saturation} up to at least $A/Z = 56.$ 
The question to address is then, whether the injection rate saturates and vanishes with growing $A/Z$ or,  
on the contrary, the accelerated  \emph{protons} generate such strong and long waves and/or magnetized eddies downstream  
that they scatter and inject species with $A/Z\gg 1$ more efficiently than (``reductio ad absurdum``) the \emph{protons}  themselves. 
We stand by the statement, that the unlimited growth of the selection rate with $A/Z$ is unphysical or, at a minimum, imposes quite unusual 
constraints on the scattering turbulence.
It is worth mentioning, that the chemical enhancement of heavier elements with $A/Z>8$ in \cite{CaprioliPRL17} is 
determined in the upstream plasma, because at time of measurement these ion species have not yet developed the universal downstream DSA spectrum. 
Whether this approach is justified at first place is controversal. The question of the exact position of maximum of $\eta_{\mathrm{sel}}$ as function of $A/Z$ is debatable 
and there is indeed no consensus yet.
Physically, its position should also depend on the current maximum energy of protons since resonant waves produced by them may scatter 
particles with larger $A/Z.$ But the particle scattering rate in general decays with the growing wave length, so this effect should not be overestimated.

CR abundances of heavier elements, such as iron, have a weaker comparative potential for the verification of the $A/Z$ scaling 
of injection efficiency than the rigidity spectra of $p$/He, $p$/O, $p$/C used in this paper. The reasons are of two kinds; first, there are not 
yet rigidity spectra of the ratios of heavier elements comparable in quality to the recently published AMS-02 data for the above ratios. The integrated 
abundancies are available, but they are affected by many factors either unrelated to the microphysics of injection selectivity in collisionless shocks or highly uncertain. 
These include, but are not limited to, CR spallation effects during the propagation to the Earth, the possible contribution from the disintegration of dust grains, 
and the uncertainty in the ionization state during the injection process.

\subsection{Do 2D simulations produce more credible results?}
\label{sec:2Dvs1D}
Although most fundamental aspects of shocks are one-dimensional, there are indeed essential phenomena that cannot be fully 
understood if two or even just one coordinate is ignored. Obviously, realism demands to trade the fully 3D simulations for  
an adequate resolution. In the case of hybrid simulations this requirement concerns both the particle statistic and parameters of the numerical 
grid. Despite the progress in computational performance, the 3D simulations meeting such conditions are hardly possible now, therefore
2D modeling appears as a computationally expensive, but a plausible compromise. With this it is tacitly assumed that 2D simulations of 
collisionless shock particle acceleration produce more credible results than 1D simulations do.   
However, there are well-known aspects of particularly the 2D fluid \cite{Kraichnan2Dcasc1967PhFl...10.1417K,Weiss2DturbCasc1991PhyD...48..273W,Biskamp2Dturb89},
absent in 3D, that makes the preponderance of 2D over 1D modeling questionable for the studies of particle scattering and acceleration. 

First, an
inverse cascade in 2D fluid leads to coherent structures that may
become responsible for an excessive particle scattering and reflection,
i.e. \emph{injection}. Although the difference between 2D and 3D dynamics
is not so explicit in the MHD, the conditions and the character of
an inverse cascade in 3D MHD are not so robust as in the 2D case \citep{Pouquet76}.

Second, high computational demands of injection studies force to elongate 
the simulation box significantly in the shock normal direction
which is unnatural for a shock alignment along its front.
The small transverse box size 
renders the 2D simulations quasi one-dimensional, but an artificial
scale introduced by it can cause an artificial 
periodicity for ions with large Larmor radii and 
is likely to determine the size of the scattering structures \citep{NganQuasi2DTurb,CelaniQuasi2DTurb}, shock corrugation scale, and
the inverse cascade anisotropy. Structures that appear to be strongly
influenced by the box geometry are seen in some advanced 2D simulations,
e.g., \citep{CaprioliFilam2013ApJ...765L..20C}. They would, perhaps,
be acceptable for a shock tube setting, but problematic for a freely
propagating shock front.

Third, besides their exaggerated magnetic strength, coherent structures in
the 2D downstream turbulence are highly consequential for the particle
injection for another reason as well: as the particle motion is considered
three-dimensional, these structures, being extended along the ignorable
coordinate, dramatically increase the effective scattering cross-section
for particles. 
The particles are scattered by these structures regardless
of their velocity projection on the ignorable coordinate. In a square
box of size $L^{2}$ and the typical scale of the scattering structure
$\sim a$, this enhancement is a factor of $L/a\gg1$, compared to
the 3D box $L^{3}$. 
Such particle dynamics may indeed result in an
excessive return upstream of particles with high $A/Z$, that would
in 3D pass through these structures, not to mention the uncertainty
of their formation in the 3D.

\subsection{Importance of the He-driven waves}\label{sec:He-driven}
\begin{figure*}
  \centering
  \includegraphics[width=0.49\textwidth]{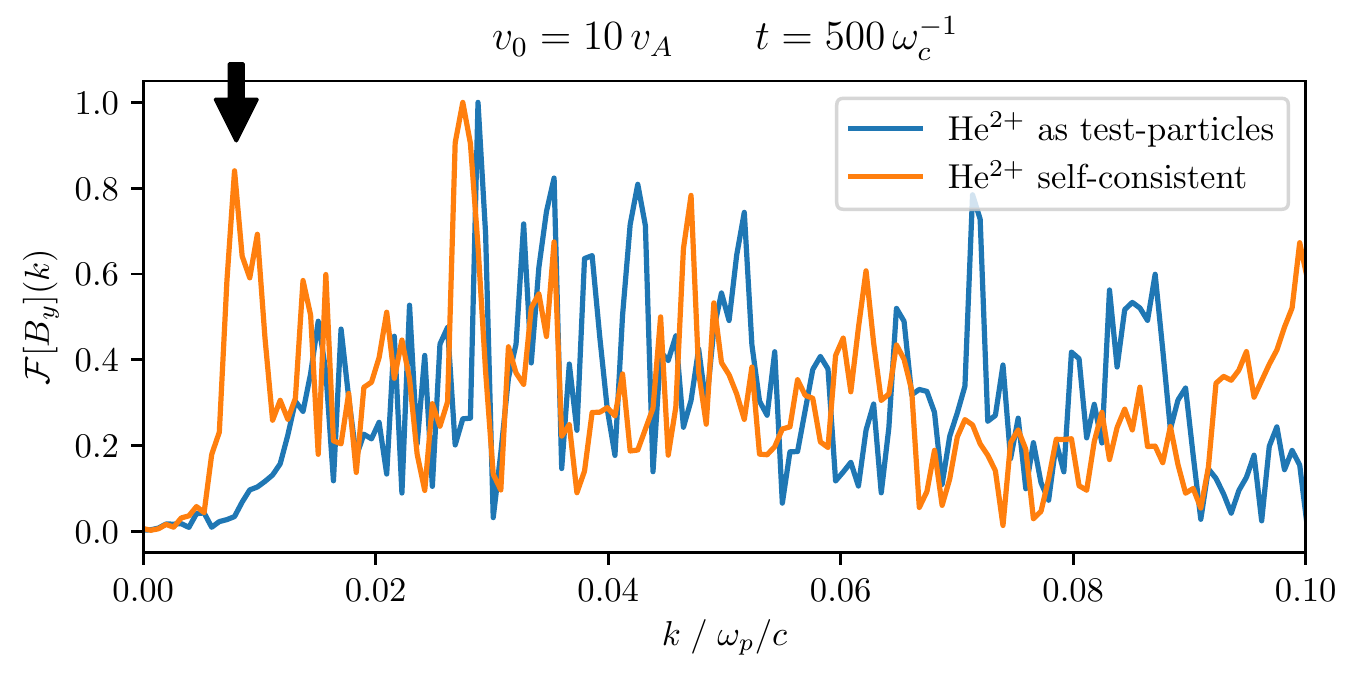}
  \includegraphics[width=0.49\textwidth]{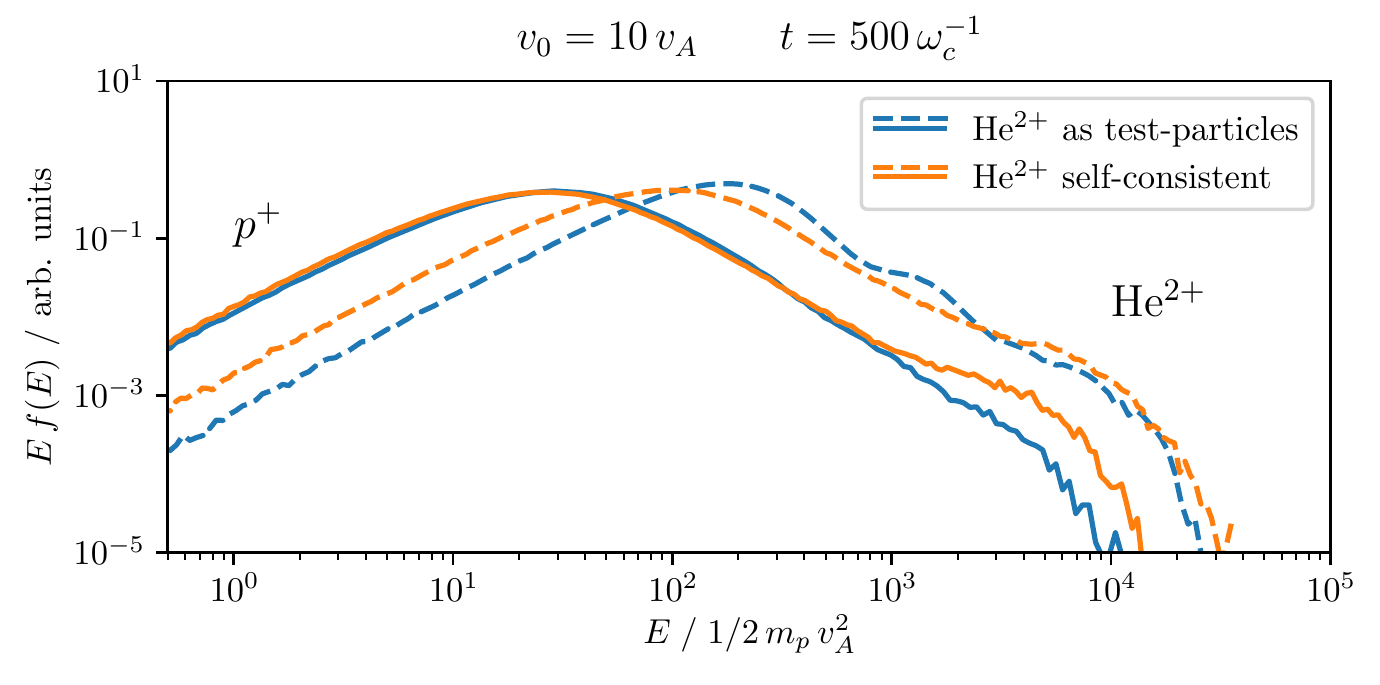}
  \caption{ (left) Spectra of the magnetic field $\mathcal{F}[B_y](k)$ in the whole simulation box for simulations with 
  a single-ion species (only protons self-consistent, He ions as test-particles) (blue) and a two-ion species (90\% protons and 10\% He ions) (orange) plasma.  
If He$^{2+}$ population is present, a component at lower $k$ appears in the spectra.
(right) Downstream energy spectra of protons (solid line) and He$^{2+}$ ions (dashed line) 
for simulations with a single species (100\% protons) (blue) and a two-ion species (90\% protons and 10\% He ions) (orange) 
plasma. In the former case the He ions are treated as test-particles, that move in the fields, created by protons.
The proton spectrum (solid line) differs only in the high energy part. 
Due to the longer waves, that are generated if helium is included self-consistently, 
more protons are accelerated to higher energies. For the energy spectrum of helium (dashed line) the main difference 
at $t=500 \omega_c^{-1}$ is close to the cut-off, where some enhancement is evident and in the thermal part of the spectrum.}
  \label{fig:he-influence}
\end{figure*}

Besides already stated, the peculiarity of our study is in the interaction of the shock with a large number of different species for 
which adequate particle statistics is vital, especially when exploring the high-energy tails of the distribution functions. 
As we briefly discussed in the Introduction, Sec.~\ref{sec:intro}, the He-driven waves significantly enrich the wave spectrum 
by their parametric interaction with the proton-driven waves, thus facilitating particle thermalization downstream.
Because of the high computational demand in the 2D hybrid numerical studies of injection, the heavier ions, including 
He, are either treated as test-particles, or included quasi-self-consistently with low statistic and an extremely low abundance. The latter protects the hybrid simulations from an 
excessive numerical heating, which otherwise would unavoidably represent a serious problem, but at the same time completely excludes the generation of the He-driven waves, by 
making He-component dynamically unimportant. 
In Fig.~\ref{fig:he-influence}a the spectrum of the transverse magnetic field $\mathcal{F}[B_y](k)$ in a two-ion species (90\% protons and 10\% He$^{2+}$) 
plasma is shown for $t=500\omega_c^{-1}$ in comparison with the corresponding spectrum in a pure hydrogen plasma.  If the abundance of He$^{2+}$ component is high, a component at lower wave-number
$k$ appears in the spectrum (shown by an arrow). The critical role of the self-consistent, as opposed to test-particle, treatment of He$^{2+}$ population is confirmed by 
the enhanced number of downstream protons with high energies and an increase in the number of helium ions near the cut-off (Fig.~\ref{fig:he-influence}b). 
In general, DSA is a bootstrap process, in which the particles with high energy drive longest waves that help to accelerate them. Here helium paves the way for protons. 
The well established part of the He spectrum is 
in turn dominated by more abundant protons in both cases.

\section{Summary}

We investigate the particle injection into the DSA using self-consistent hybrid simulations. 
We provide sufficiently detailed Mach number scans of the $p$/He injection \emph{ratio}, that is needed to test the injection bias. 
 It should be emphasized that the rigidity spectra of the \emph{fractions} of different species do not depend on the relation, in which 
these elements are in the most productive SNRs, as protons are considered to be dynamically most important species.
The reduced spatial dimensionality of the simulations allows us to increase the particle statistics and grid resolution dramatically. 
Our simulations show that  selection rate of different ion species increases with $A/Z,$ 
saturates, and peaks as a function of Mach number. They correctly predict the decrease in proton-to-helium
ratio with increasing rigidity, Fig.~\ref{fig:pHe-ratio},
at almost exactly the rate $\Delta q\approx0.1,$ measured in the experiments
for $\mathcal{R}\gtrsim10$ GV. 
At lower rigidities, 
the difference between the data and our predictions is significant.
Based on the discussion made in the introduction,  
the difference \emph{must} occur because the equations of motion, 
Eqs.~(\ref{eq:RigMotion}), (\ref{eq:CoordMotion}), 
for protons and helium ions deviate toward lower rigidities.
The most likely cause of this deviation is particle interaction with the 
turbulent solar wind in the Heliosphere, but the interaction with the ISM 
turbulence may also contribute, again, because the equations of motion are 
different for $p$ and He in the low-rigidity range. By contrast, the deviation 
from the AMS-02 data in the high-rigidity range, where the equations of motion 
for $p$ and He become identical, is insignificant as expected.  This deviation 
is much less than the difference $\Delta q \approx 0.1$. Whether it comes from a 
simplified integration over the SNR in Eqs.~(\ref{eq:number}) and (\ref{eq:spectra}) 
or 
it is a mixing effect from different SNRs or spallation in the ISM, 
remains unclear. The difference is small enough to be accounted for by any of these phenomena. 
Except for this uncertainty, the suggested mechanism for $A/Z$-dependence of the injection fully explains the measured $p$/He 
ratio. 
Our interpretation of the elemental ''anomaly`` is therefore intrinsic to collisionless shock
mechanisms and does not require additional assumptions, such as the contributions from several
different SNRs, their inhomogeneous environments or acceleration from grains.

\acknowledgments
The research was supported by the DFG within the Research grant 278305671, by 
RSF 16-11-10028, and by NASA ATP-program within grants NNX14AH36G and 80NSSC17K0255.
Simulations were performed using the computing
resources granted by the North-German Supercomputing Alliance (HLRN) under the project mvp00015.

\appendix
\renewcommand{\thesubsection}{\Alph{subsection}}
In hybrid modelling the evolution of the ion distribution function $\displaystyle f$ is governed by the kinetic Vlasov equation:
\begin{equation}
    \frac{\partial}{\partial t} \, f + \mb{v}_i \nabla \, f 
	+ \frac{q_i}{m_i} \left( \mb{E} + \frac{1}{c} \, \mb{v}_i \times \mb{B} - \eta\, \mb{J} \right) \frac{\partial}{\partial v} \, f = 0.
	\label{eq:Vlasov}
\end{equation}
Here $\mb{E}$ and $\mb{B}$ are the electric and magnetic fields and
 $\mb{J} = e (n_i \, \mb{v_i} - n_e \, \mb{v_e}) \approx e \, n (\mb{v_i} -\mb{v_e})$ 
is the current density. Furthermore, $q_i = Z \,e$ and $m_i = A\, m_p$ are the ion charge and mass and $\eta$ denotes a scalar resistivity. 

The plasma electrons on the other hand are treated as a charge neutralizing massless fluid,
\begin{equation}
    n_e\,m\, \frac{d \mb{v_e}}{dt} = 0 
    = -e\, n_e  \left( \mb{E} + \frac{1}{c} \, \mb{v} \times \mb{B} \right) - \nabla p_e + e \, n_e \, \eta\, \mb{J}.
	\label{eq:electron-momentum}
\end{equation}
Here $p_e$ denotes the electron pressure, which can be calculated as $p_e = n\, k_B\, T$. In order to
close the set of equations we assume an adiabatic equation of state
\begin{equation}
    \frac{T_e}{T_0} = \left( \frac{n_e}{n_0} \right)^{\gamma-1} \qquad \text{with:} \quad \gamma = \frac{5}{3}
\end{equation}
We found that the use a polytropic equation of state, as in e.g. \citep{Caprioli2015}, instead of an adiabatic one
does not change the energy spectra significantly and  the injection rate behavior for large $A/Z$ is not affected by the prescription for the electron equation of state.
Furthermore, an effective adiabatic index, based on the assumption
of equilibration of electron and ion temperatures in the downstream
might be justified for weak shocks, as \citep{Vink2015} predicts a thermalization
between electrons and ions at low M shocks.
However, at higher Mach numbers ($5 < M_s < 60$) the same authors predicted
a behavior of $T_e/T_i \propto M_s^{-2}$. Hence the assumption of electron 
and ion temperature equilibration might not hold.

Furthermore, we use the low-frequency magnetostatic model, neglecting the displacement current
in the Ampere's law.

\subsection{Implementation}

The particle-in-cell method is used for the ion plasma-components.
A first order weighting is applied to interpolate the fields to the particle position, 
as well as to obtain ion current and charge density from the known positions of the ion-particles relative
to the grid points.
The \emph{Boris-algorithm} \citep{Boris1970} is used to update the particle positions and velocities.
The magnetic field evolves according to Faraday's law, 
while the electric field is calculated from the electron momentum equation, Eq.(\ref{eq:electron-momentum}).
The equations of the evolution of the fields are discretized using second order finite difference stencils:
\begin{eqnarray}
    \mb{B}^{n+1/2} & = & \textstyle \mb{B}^n - \frac{\Delta t}{2} \nabla \times \mb{E}^n\nonumber\\
    \mb{E}^{n+1/2} & = & \textstyle F(\mb{B}^{n+1/2}, n_i^{n+1/2}, \mb{J}_i^{n+1/2})\nonumber\\
    \mb{B}^{n+1}   & = & \textstyle \mb{B}^{n+1/2} - \frac{\Delta t}{2} \nabla \times \mb{E}^{n+1/2}\nonumber
\end{eqnarray}

Technically, the problem of the time evolution of the fields reduces the problem of calculating $\mb{E}^{n+1},$ which cannot be calculated directly, since $\mb{J}_i^{n+1}$ is not known. 
Several different methods have been proposed in the literature, see \citep{Winske2003} for a review. 
We use a predictor-corrector method, which is simple and has good energy conserving properties.
The algorithm has the following steps:

\begin{enumerate}
    \item advance $\mb{B}^n$ and $\mb{E}^n$ to the time step $n+1/2$
    \begin{eqnarray}
        \mb{B}^{n+1/2} &=& \textstyle \mb{B}^n - \frac{\Delta t}{2} \nabla \times \mb{E}^n \nonumber\\
        \mb{E}^{n+1/2} &=& \textstyle F(\mb{B}^{n+1/2}, n_i^{n+1/2}, \mb{J}_i^{n+1/2}) \nonumber
    \end{eqnarray}
    \item predict fields $\mb{B'}^{n+1}$ and $\mb{E'}^{n+1}$
    \begin{eqnarray}
        \mb{E'}^{n+1} &=& \textstyle 2\, \mb{E}^{n+1/2} - \mb{E}^{n} \nonumber\\
        \mb{B'}^{n+1} &=& \textstyle {B}^{n+1/2} - \frac{\Delta t}{2} \nabla \times \mb{E'}^{n+1} \nonumber
    \end{eqnarray}
    \item advance particles using the predicted fields 
    and obtain $n_i'^{n+3/2}$ and $\mb{J'}_p^{n+3/2}$
    \item calculate predicted fields $\mb{B'}^{n+3/2}$ and $\mb{E'}^{n+3/2}$
    \begin{eqnarray}
        \mb{B'}^{n+3/2} &=& \textstyle \mb{B'}^{n+1} - \frac{\Delta t}{2} \nabla \times \mb{E'}^{n+1} \nonumber\\
        \mb{E'}^{n+3/2} &=& \textstyle F(\mb{B'}^{n+3/2}, n_i'^{n+3/2}, \mb{J'}_i^{n+3/2}) \nonumber
    \end{eqnarray}
    %
    \item determine $\mb{B}^{n+1}$ and $\mb{E}^{n+1}$
    \begin{eqnarray}
        \mb{E}^{n+1} &=& \textstyle \frac{1}{2}\, (\mb{E'}^{n+3/2} - \mb{E}^{n+1/2}) \nonumber\\
        \mb{B}^{n+1} &=& \textstyle {B}^{n+1/2} - \frac{\Delta t}{2} \nabla \times \mb{E}^{n+1}. \nonumber
    \end{eqnarray}
    %
\end{enumerate}

\bibliography{Literature}

\end{document}